# A Table-Binning Approach for Visualizing the Past

Nicolas Turenne,
Univ. Paris-Est, LISIS, INRA, nturenne.inra@yahoo.fr

**Abstract**— Large amounts of data are available due to low-cost and high-capacity data storage equipments. We propose a data exploration/visualization method for tabular multi-dimensional, time-varying datasets to present selected items in their global context. The approach is simple and uses a rank-based visualization and a pattern matching functionality based on temporal profiles. Ranking categories can be specified in a flexible way and are used instead of actual values (value reduction into bins) and plotting it over time in an unevenly quantized representation. Patterns that emerge are matched against a set of eight predefined temporal profiles. The graphical summarization of large-scale temporal data is proposed and applicability is tested qualitatively on about eight data sets and the approach is compared to classic line plots and SAX representation.

## INTRODUCTION

As processor speed does not progress as fast as storage capacity two issues may concern their data processing. Typical features of such data are of two kinds, time and high-dimensionality [1]. Usually, classical approaches try to apply dimension reduction methods (e.g., principle component analysis) to derive major temporal trends [2][3][4].

In this paper, we present a method, Time Rank Levels, to visualize such sets of variables having at least a numerical (ordinal) parameter changing over time. We argue that identifying an object among the global surrounding set of items can help to understand its relative position with the help of its context. A zooming effect plays a key role about this property. A second key role to enrich a zoom effect is sorting such a given object according the context of all other objects. In this way, for our technique we adopt a Focus+Context-like data exploration in the same way developed by a table lens approach to look tabular data [5]. Another key point is explicit time visualization. As data will be represented as large tables of items without knowing intern relations, we adopt a visual representation close to polylines in parallel coordinates plot [6] to show concretely at which time-point an object occurs comparing to occurrences of other time-points: "after" and "before" relations. In summary, using Focus+Context technique and parallel axes representation of time, our technique aims at visualizing explicitly a given object explicitly over time on parallel axes (i.e. columns): -a) without requiring data deletion (all data point can be localized) even if we reduce the amount of possible values -b) showing "after/before" relations, and focus technique and -c) having relative context with a condense form ranks of values instead of attribute values about all other objects surrounding. A challenge about visual analytics has been launched recently [7]. For this dataset social network analysis is asked; our technique could analyze suspicious activities but no metrics and train/test information is available about this. Nevertheless, except this challenge, some datasets including a time parameter are freely available and we used them to test efficiency of our technique to re "discover" pieces of knowledge published about the data.

In Section 2, we introduce the related work. Section 3 presents the Time Rank Levels approach. Section 4 gives examples of applicability.

## 1 RELATED WORK

In this section, we recall how time is displayed in maps or graphs.

### 1.1 Explicit Time-Dependant Information

Of the seven "visual variables" of Bertin (1983) [8], most approaches to time-dependant visualizations use position to represent time. Most commonly, color, as value (dark/light), form and size are used to represent the temporal data. Color (its component value) is frequently used for visualizing cyclical patterns (weekends, months, hours of darkness, etc.). Form tends to be used for qualities which do not change over time, such as categories. Another widespread way to represent time-stamped points uses time as any other variable. Data are represented against a time flow along one axis, meaning that representations summarise key information about that flow on this axis, i.e. a line graph or a "time bar" [9][10].

But this graphical choice does not assume any specification about the dimensionality of the space to represent. [11] describe a symbolic representation (SAX: symbolic aggregate approximation) of data to produce a piecewise vectorization capable to make good comparison of subsequences times series and find clusters of similar series or common patterns occurring at same time ranges between series. Such a piecewise vectorization makes a kind of smoothing in data very close to our value reduction in our approach. We will compare this representation to our in next section.

## 1.2 Multidimensional Information

In our purpose, we manage a data matrix where each line points out to a variable (a dimension) and columns mean timed values of a parameter. Such a matrix is generally huge and contains several thousands, even millions of lines. In multivariate data analysis, the correspondence analysis is a successful approach to feature reduction and plotting a small number of dimensions. Such an approach summarizes tendencies within principal components but does not make a user to interact with specific features. When a data matrix is quite sparse, a crude display of data, even with re-ordering involves an interesting technique for clusters identification. But in our case the matrix contains values almost for all variables almost for all time-points. An interesting technique is called the parallel coordinate technique [6][12]. Each observation in a dataset is represented as an unbroken series of line segments which intersect vertical axes, each scaled to a different variable. The value of the variable for each observation is plotted along each axis relative to the minimum and maximum values of the variable for all observations; points are then connected using line segments. Each data item is presented as a polygonal line, intersecting each of the axes at that point which corresponds to the value of the considered dimensions. Parallel coordinate has not been used to represent time because axes are independent practically but it should be interesting for a few time points, in other cases axis are not readable. A multiscale technique [13] can be fruitful to change the level at which distance between points can be viewed. In general, human interaction is widely practiced to perform temporal data exploration [14][15]. This technique is in so far related to human interaction with by visual navigation on the axis [16].

A concurrent way to multiscale but without implying graphical interaction is a pure pixelization of data [17]. Another approach of dense pixel techniques is to map each dimension value to a colored pixel and group the pixels belonging to each dimension into adjacent areas [18]. Since in general dense pixel displays use one pixel per data value, the techniques allow the visualization of the largest amount of data possible on current displays (up to about 1,000,000 data values). If each data value is represented by one pixel, the main question is how to arrange pixels on the screen. Dense pixel techniques use different arrangements for different purposes. By arranging the pixels in an appropriate way, the resulting visualization provides detailed information on local correlations, dependencies, and hot spots. Using a zooming effect but also taking benefit or sorting property the table lens component is ideal for visualizing large tabular datasets [5]. In an interactive way clicking on a row expands the entire row and displays the value of the row in a textual form. Line Graph Explorer approach by clicking on a column header causes the entire table to sort based on that value [19]. But in our approach sorting is done independently from one column to another. Such an approach has also been developed for unstructured data, in the Document Lens system [20] documents are displayed as small pictures of pages layed out in a grid in the reading direction. Hence, in such overview the different documents are not readable but only their visual aspect as zoomed in pictures can be an index to drive the user to a desired document. As other systems, Document Lens uses greeking to display text in small size. This technique aims at representing each line with a part of line having a length in proportion to the line number of characters. Practice of greeking is quite used in information visualization leading to chunk of document recognition only by patterns and shape of typical parts of documents.

## 1.3 Ranked Information

A traditional way to overview ranked information is ordering data to make a top-10 or top-100 of information. Origin of such data organization is certainly very old because practical, intuitive and easy to understand for comparison. The original Olympic Games, racing, first recorded in 776 BC in Olympia, Greece, are a kind of primitive ranking of social performance. More recently in computer gaming, for a small set of variables, rank-time relation has been exploited to enlighten classification of success for player role [21]. To get a sense of the fastest levelling times, it computes the top 1st percentile, 5th percentile, and 10th percentile of times it takes to get from one level to the next. Seo *et.al.* (2005) [22]show that data clustering of large sets can be supported by ranking data according a specific parameter. Closely to previous techniques of multidimensional visualization, Andrienko *et.al.* (2001) [23] studied impact of scales transformation (straightening of lines, normalization with a statistical parameter) of parallel coordinates and show that transformation can be useful for a specific use.

## 1.4 Time Ontology Usage

Use of expert knowledge to describe information received high attention from computational communities and applications areas of time-dependent data processing (industry, semantic web…) [24][25]. A big effort has been developed to make possible manual, and perhaps automatic, annotations of documents with an ontology of time. These kinds of studies largely deal with calendar forms of time description. In numerical data mining and time series very few studies try to integrate symbolic time ontology. [26] discuss a new technique of time series analysis based on moving approximation transform mapping time series values in tendency domain to replaces time series by the sequence of "local trends" and definition of perceptions like *"Quickly Increasing", "Very Slowly Decreasing"* etc. Our approach tries to take into account shape descriptions of interest.

For sure it should depend upon domain of application, and at present our approach makes a visual mapping between a perception-like time ontology with our graphical data representation.

## 2 TIME RANK LEVELS

### 2.1 Basic Assumptions

Time Rank Levels is a technique for visualizing evolving multidimensional information according a timeline. We define the variables and the parameter as follows. Let t be the dimension of time. The domain of t is discrete and takes n time points, $t \in \{1,...,n\}$. n is not necessarily high. Let E be a descriptive space composed of p variables. $E = \{e_1,...,e_p\}$. p can be very high. Let q being a quantitative parameter, $q \in \mathbb{R}$. Each variable $e_j$ is defined over the q domain: $e_j(t) = q_j(t)$.

Hence E becomes $E(t) = \{q_1(t),...,q_p(t)\}$, the question is how to visualize E(t) ?

points are presented in a global plot and keep "time's arrow" metaphor of linear progress.

It is related to the fact that a) scale is a mean to smooth tendencies, b) parallel axis (i.e. columns) make relations between events and c) ranking is ordering processing to extract relevance of an event. Firstly we make assumption that each time point can be seen as a separate coordinate and can lead to a specific axis in a parallel axis visualization. Secondly we reduce size scale of data on screen so as to visualize the whole set of data for a given time point. Thirdly we sort data according to q numerical parameter such that for any couple of variable $(e_i, e_j) \in E^2$ for a given time point $t_k$, $e_i$ is higher that $e_j$ on the $t_k$ column if $q_i(t_k) > q_j(t_k)$.

In most cases equality can occur for larges sets. Because location at a level on the column is dedicated to a unique variable, if lots of variables get a same ranking it can distort the relative position of a variable compared to another leading interpretation that it is less ranked or better ranked. To avoid such distortion and misinterpretation in visualization, a binning is necessary to merge some variables at an equivalent place on the corresponding column.

Finally to make relative importance of list of items having same value we plot data on boxes covering large ranges of same values items. Fig.1. gives an example transforming data (from I to IV) so as to visualize data for a unique time point.

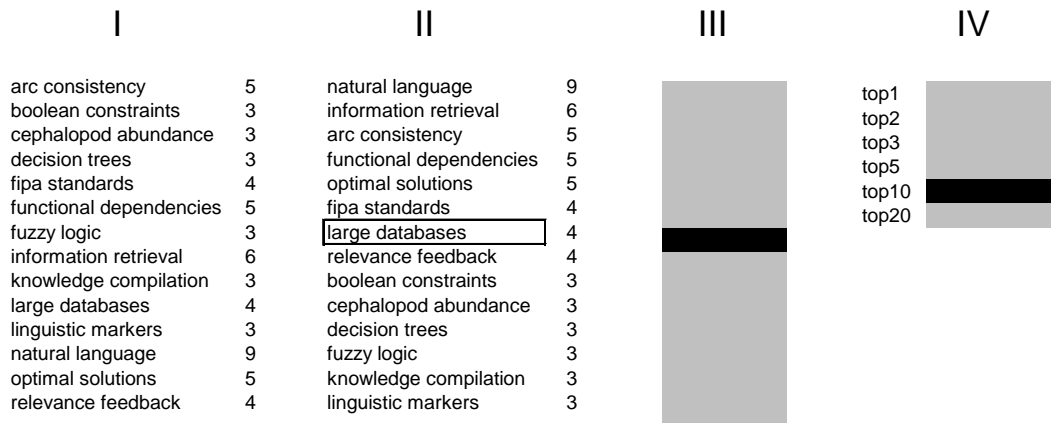

Fig. 1 Data transformation. From left alphabetical order, to right binning ranking..An item is chosen and blackened.

Time Rank Levels approach tries to answer this issue in a simple way. The basic idea is to take $t_i$ as an axis, visually represented as a distinct column; so $E(t_i) = \{q_1(t_i),...,q_p(t_i)\}$ is plotted according column $t_i$. The t domain is sorted in an ascending order. Therefore columns are contiguously presented from left (old time points) to right (recent time points). All time

In this example items are linguistic phrases meaning topics in computer science. The parameter is the number of occurrences in documents. The first set of features (Fig.1. I) represents a rough list sorted alphabetically. Hence we transform this set into a second one resulting from a decreasing sort according parameter values (Fig.1. II). At this step we can choose a specific item (SI), for instance "large databases", to localize it in the items list overview (i.e. table lens). A SI could be any item of the overall set of items in the dataset. A following transformation makes SI in a black-colored box and all other in gray-colored box (Fig.1. III). SI position is clearly identifiable because list size is not long but in case of large sets localization is not easy. To solve this problem we put the SI into a box knowing its position. In our example the SI rank is $7^{th}$, hence it cannot be located at least in top-10 of items. The final table is the set IV (Fig.1. IV) where top-10 is blackened and all others are grayed out. This example describes one table corresponding to one time point, in the case of several time points, all tables are juxtaposed from left (early time point) to the right (late time point) as it looks like on Fig. 2. Notice the value reduction is the same for each timepoint but ranking is repeated independently for each period of time.

Without transformation in top boxes a rough map should be seen as in on Fig.3. A given item is highlighted to see the evolution of its position compared to other items and related to the magnitude of its parameter. In this example we see the Gross Domestic Product evolution per year of Austria for the last 15 years. But sometimes we fall on an ordering bias since lots of variable can take same values and lead to a fluctuation of a variable position at a given time point. Hence the fluctuation points out to the same meaning and induce a misinterpretation of the curve pattern. To correct this bias we can reduce a block of variable into interval of parameter as we described in our example (Fig. 4). For instance at the first line of the rough map, an item occurs at the top because its parameter value is the highest. At present the size of each interval is build according to a range of ranks but blocks could be made with ranges of parameter values using some knowledge about the data domain. In a generic way the first lines concerns a few range of ranks and the last lines can contains some thousands of ranks because low values can be more common than highest ones in the data. Interval of ranks leads to a smoothing of the time-dependent curve and its interest is maximized in the case of a high dimensional variable space, in other case the smoothing crushes data as seen in Fig. 5.

Experiments show that ranges of ranking become useful using more than 300 variables. Results can be visualized with datasets in section 4.

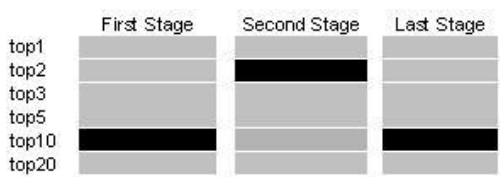
Fig. 2. Time Rank Levels with a given item blackened.

### 2.2 Building visualization

To achieve a plot with Time Rank Levels, 5 steps are required. First step is formatting of data, the second, third and fourth steps are processing of data, last step is interaction step. As preliminary processing it is necessary to apply a data reduction strategy which may depends upon knowledge to fix breakpoints :

1- Let define couples (upper limit of rank, step) to split the range of new dimensions
2- Sort new dimensions from step = 1.

For instance in our GDP database containing 191 items (country names), we define couples as follows (20,1), (100, 5) and (191, 10). We hence obtain the following new dimensions (top-1, top-2, ... , top-19, top-20, top-25, top-30, … , top-95, top-100, top-110, top-120, … , top-200). Let suppose a country name get rank 122 it will be classified into the dimension top-130. Recall the set of new dimensions is unique and specific for a whole given dataset but ranking will be achieved for each timepoint of the dataset.

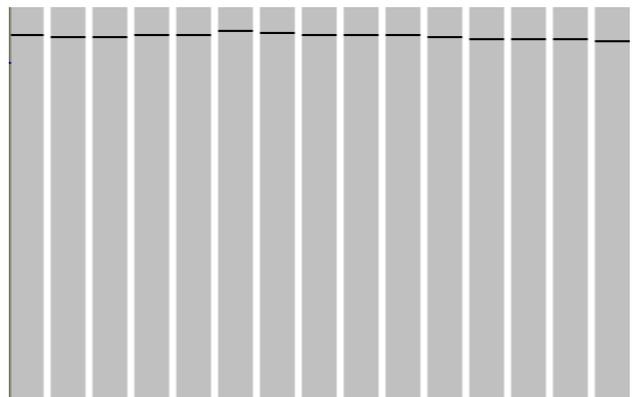
Fig. 3. Time Rank Levels with unbinning visualization.

Once new dimensions are known, following steps will implement visualization :

Step 1- Requiring format is column-based; meaning that each column represents a list of items present during the current time point. The timeline is from left to right as standard representation of time evolution in a two-dimension map. Hence the extreme left column is the oldest. Each column contain an item for each line and adjacently a numerical (integer of real) value of a same parameter for all items.

Step 2 - Each column is decreasingly sorted from top to bottom according to value parameter.

Step 3 - Assign data to new dimensions (in case where items list size is greater than 100 items).

Step 4 - Reduce scale of boxes about 20% (30 x 5 pixels).

Step 5 - Select a specific item. Identify box it belongs for each time point and blacken them, gray all other boxes.

Step 6 – Compare the specific item profile to a shape pattern in time ontology.

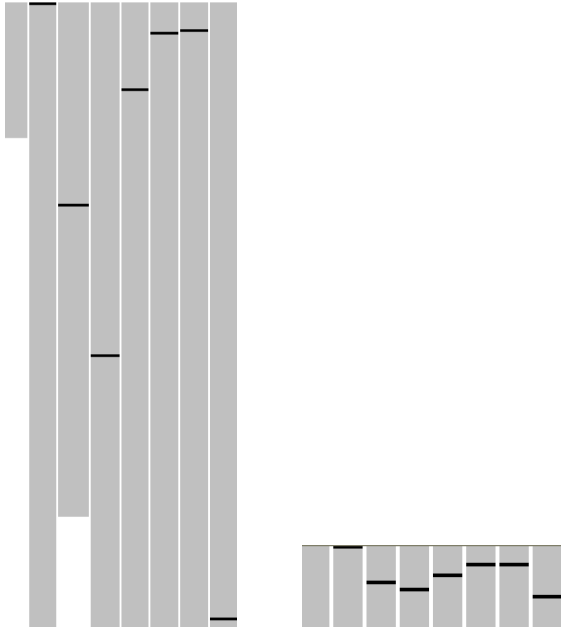

(a)      (b)
Fig. 4 Position fluctuation (a) unbinning map (b) binning map.

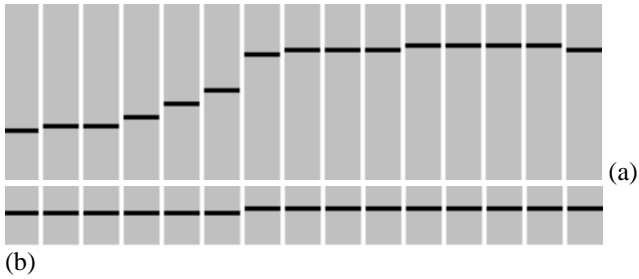

(a)

(b)
Fig. 5 Range effect (small-sets) (a) unbinning map (b) binning map.

In the method the user interacts by choosing a variable to make its profile appear.

### 2.3 Ontology of temporal profile

Even if representation, shown in fig.2, handles thousands of dimensions, it is possible to exploit easily two dimensions for interpretation; interpretation is required to make representation useful. Ontologies of temporal profiles already exists in 2-dimensions to help interpretation. It is related and permitted by keeping idea of "time's arrow".

We are specifically interested in timeline representation able to capture all information of variables. In such framework classical interpretation of shapes are based upon slopes and breakpoints. For evaluation and for classification of a time profile we derive eight classical temporal patterns/profiles. The different temporal patterns/profiles depend on the selected time frame. Our patterns are chosen in a time frame having a 2-D timeline with continuous occurrence. To understand changes over time, plotting has also been used as a natural tool to show how events can occur (look at §II.A) and some useful patterns were commonly considered as gold standards: Decreasing trend, Cyclic, Normal, Upward shift, Increasing trend, Downward shift.

Table 1 temporal profiles

| Maps | Labels |
|---|---|
| | SPIKE |
| | FLUTTERING (MULTISPIKE) |
| | PROGRESSIVE DECREASING |
| | PROGRESSIVE INCREASING |
| | MULTISTAGNANT |
| | LATE MONOSTAGNANT |
| | EARLY MONOSTAGNANT |
| | EMERGING |

Close to curve shape showing specific evolution of a phenomenon (§II.B) we made a classification of curve shapes for the Time Rank Levels technique. Eight shapes can facilitate interpretation of evolution of a given highlighted item (Table I). There are probably other interesting shapes. We define a pattern (i.e. a shape) as follows. Let $P(e_i) = \{q_i(t_1),...,q_i(t_n)\}$ be the profile for $e_i$ variable, $<e_i> = \frac{1}{n}\sum_{j=1}^{n} q_i(t_j)$ is the average parameter value of $e_i$, and $V_i(t_j) = \{\exists t_j, q_i(t_j) \approx q_i(t_j+1) \approx q_i(t_j+2)\}$ be a plateau consisting at least of three contiguous values of equivalent magnitude, $\lambda$ is a threshold value of q, $\varepsilon$ is a threshold value for a variation of q. 'Spike shape' shows a level higher than the medium level over the sets of intervals $\{\exists t_a \forall t_y \; q_i(t_a) >> q_i(t_y) \mid q_i \in P(e_i)\}$. 'Fluttering shape' shows alternating high and low levels and at least a saddle point $\{\exists t_a \exists t_b \forall t_y \; q_i(t_a), q_i(t_b) >> q_i(t_y) \mid q_i \in P(e_i)\}$.

'Progressive Decreasing shape' makes the levels decreasing from left to right $\{\forall t_a \forall t_b, t_a > t_b \text{ and } (q_i(t_a) - q_i(t_b)) > \varepsilon \mid q_i \in P(e_i)\}$. 'Progressive Increasing shape', reversely, makes the levels increasing from left to right $\{\forall t_a \forall t_b, t_a > t_b \text{ and } (q_i(t_a) - q_i(t_b)) < \varepsilon \mid q_i \in P(e_i)\}$. 'Multistagnant shape' corresponds to two or more plateau, each one consisting of consecutive levels $\{\exists t_a \exists t_b, V(t_a), V(t_b) \mid q_i \in P(e_i)\}$.

We consider that two levels can build a plateau a difference of one range of ranking is acceptable to merge levels into a plateau. 'Late Monostagnant shape' consists of a plateau occurring at the half part of the map on the right $\{\exists t_a, t_a > t_{n/2} \text{ and } V(t_a) \mid q_i \in P(e_i)\}$. 'Early Monostagnant shape' consists of a plateau occurring at the half part of the map on the left $\{\exists t_a, t_a < t_{n/2} \text{ and } V(t_a) \mid q_i \in P(e_i)\}$. Finally the 'Emerging shape' is less typical in term of its shape but concerns a curve having a major part of levels over a threshold line to underline the emerging property of the given item $\{\exists t_y \; q_i(t_y) > \lambda \mid q_i \in P(e_i)\}$. We need to precise that $q_i$ could have null values and so $e_i$ will not systematically appear in each column. And profiles are used for the whole dataset. These patterns are not mutually exclusive, or could a timeline exhibit multiple patterns simultaneously; in such cases interpretation is of course less trivial.

## 3 APPLICABILITY

The following studies about temporal data and their visualization use datasets to adapt the method and export displays.

### 3.1 Random database

From a dictionary of 5,000 items labeled with a name and a parameter value between 0 and 1, we settled randomly 10 ordered sets. Each set representing a time point.

1973 items have an occurrence of their parameter higher than 0.95. We have limited user-interface actions and interpretation with a sample of 14 items among them. For them we extract their ranking levels map. In this data simulation we consider the two cases. Case 1, the value parameter is significant any time (i.e. plot without NULL items). Case 2, if in a time point a value belongs to the last range rank, we assume that the item does not occur in the current time point (i.e. plot with NULL items). In case 2, and with a sample of 14 items we get approximately an equipartition among fluttering, multistagnant, late and early monostagnant profiles.

### 3.2 Document database

Here we use the visualization for displaying items extracted from a semi-structured document archive. The corpus has been retrieved on the Science Citation Index (SCI) server (i.e. web of science) with a query consisting of cited authors from the field of text mining and concerning French publications between 2001 and 2007. The archive can be described by 1,158 documents, 2,119 citing authors, 16,055 cited authors, 11,348 terms (2,113 keywords and 9,235 phrases), 28,234 cited references, 45 countries. Phrases are extracted using a shallow parser applied on titles and abstracts.

About terms we can interpret moving topics over time as seems fashionable such as "semantic web" and "knowledge management" (progressive increasing), fields that make the topics with fashion style "conceptual graphs", "functional dependencies", "neural networks", "fuzzy sets" (fluttering), topics with more field denotation terms such as "natural language", "decision trees", "relational databases", "data analysis", "domain ontology" (multistagnant), and not new but more pregnant technical topics such as "concept lattice", "em algorithm", "clustering", "data sets", "correspondence analysis", "molecular descriptors" (late monostagnant). Some terms traduce that some technical topics get preference from community such as "machine learning", "sequential patterns", "Galois lattices", "computational complexity" (spike). (Figures in Table 1 comes from this dataset).

### 3.3 Microarray database

A microarray (or chip) is a large-scale experimental device in molecular biology to test a specific hypothesis according levels of gene expression. Here the following two-color comparative experiment [27] uses a microarray to show that Bone Morphogenetic protein 4 (BMP4), a member of the transforming growth factor-B (TGF-b) superfamily induces the differentiation of human ES cells to trophoblast. A DNA microarray demonstrates that the differentiated cells express a range of trophoblast markers and secrete placental hormones. 43,000 cDNA clones microarrays have been used to analyze genes differentially expressed in the BMP4-treated and the untreated undifferentiated H1 cells. H1 cells are melt with BMP4 and harvested at several time points for RNA extraction, amplification and DNA analysis on microarray containing cDNA clones, leading to a time series: 3 hours, 6 hours, 12 hours, 24 hours, 48 hours, 3 days, 7 days. The signal on the microarray is pointed out by two light beams. [27] have found upregulated genes at all the time points examined. Eleven genes in microarray: transcription factor. Four other genes using another technique called RT-PCR. Several hundred of genes were highly expressed at day 7. Using Time Rank Levels we find that most of genes are distributed among 'fluttering', 'early monostagnant' and 'multistagnant'. 'Early monostagnant' is clearly understandable because genes are often produced at constant-time flow. In our sample spike and fluttering are essentially due by the fact that platforms at time 24 and 48 hours are different from others; for example GRPEL1 does not occur with the same GeneBank Code at stage 24 hours. If platforms would have been the same we would probably have got more 'late monostagnant', 'multistagnant' and 'progressive increasing' profiles. We can show that CGA occurs increasingly at day 7 as in Xu et al (2002), but their analysis induced this requiring another approach than a microarray clustering method. We find some highly regularly expressed genes, such as TBD12 or GRPEL1, but different than those in Xu et al (2002) (see above) It makes think that such a visualization can supply hypothesis requiring experimental validation. We also find that SSI3 and TFAP2A are regularly expressed but not among the most expressed as in Xu et al (2002).

### 3.4 Discussion and Comparison

Comparing all datasets with random p-value < 0.05 (Fisher's test), we induce that distributions are not equivalent and specificity of profile distribution of each datasets compared to random datasets. Despite this, we showed that Time Rank Levels approach can recover some useful pattern about data and can be a general exploration approach for high dimensional temporal data. The exploration of large data sets, seen as an information flow or information stream, is an important but difficult problem. Our information visualization technique may help to solve the problem to visualizing time-oriented information. We tested validity of this approach on a random dataset and five different real-world case studies. It shows that the kind of dataset influences which kind of temporal patterns/profiles can be found. Random dataset makes a baseline to keep distribution of common random time profile. Time Rank Level approach can capture interesting emerging time profile datasets. Visualizing a large volume of data and information is a challenging task, in particular, if you are aiming to handle time-oriented information. In many contributions, complexity and structure of time (temporal aspects and dimensions) is not handled probable. Line Graph is a simple powerful mean to plot time data with a time arrow way, as we do with our table-lens method. But in our method data are smoothed by value reduction. And location of a given displayed item is visualized according the context of all others items. Line graphs are traditionally used to plot a time-dependent variable in real-time or static environment, even simultaneously some graphs (maximum a dozen). A specific implementation as Line Graph Explorer proposes high dimensionality with an overview of all items, and possibility to select a specific item with sorting property. It proposes to make comparison with other items based on a clustering functionality to group similar lines according their values profile but contrary to our method comparison is only done with a small sample of the dataset. Our approach does not require vector representation; some approaches represent time series with vector quantization such as SAX aiming at computing similarity between part or all time series. A time series is transformed into a vector representation which dimension is the size of time points of the series, and a vector dimension value is a symbolic pointer (binary or integer) corresponding to a value range knowing that a series of value ranges covers all the data. In the case of GDP database, data are described by values between $105 and $63692 and split into equal-sized areas. We find that for 8 ranges values, breakpoints are ($312, $603, $1420, $2307, $4573, $9006, $24382). "Ireland" will have the following SAX representation during the 1985-2000 date range: $IRE_{sax}(2,2,2,2,1,1,1,1,1,1,1,1,1,1,1)$. For "United-Kingdom" : $UK_{sax}$ (1,1,1,1,1,1,1,1,1,1,1,1,1,1,1), "Norway": $NOR_{sax}(1,1,1,1,1,1,1,1,1,1,1,1,1,1,1)$ and "Argentina": $ARG_{sax}(2,2,2,2,2,2,2,3,3,3,3,3,2,2,2)$ . If we computes an Euclidian-like distance called lower-bound distance used by [11] we find $dist(IRE_{sax}, ARG_{sax}) = 18,19$ and $dist(IRE_{sax}, UK_{sax}) = 7,13$. In this way $UK_{sax}$ and $NOR_{sax}$ are strictly the same though, "Norway" is more stable over time. Distances computation should point out that "Ireland" evolution is closer to "United-Kingdom" though "Argentina" GDP fluctuated in the same way with distinct periods of stability. The way for usage of such representation is similarity using a specific distance, i.e. Euclidian distance, saying if two time series are very close according their time point values pairwise. In this sense

it is an interesting way to find recurrent small patterns occurring in a time range or to detect general shape stereotypes over the whole set of time points. In our case we want to identify a rough shape such as a spike (or any other ontological shape) not time-localized. A similarity-based distance based on SAX is too constrained to achieve such detection. Our approach takes more account slope similarity, in that sense a visualization using simultaneously global context of data is a first step for such slope similarity analysis to catch similar time profiles.

## 4 CONCLUSION

Advantage of our approach is to locate and rank evolution of a single point among all other points. Some improvement of Time Rank Levels approach should be made using knowledge about the way to discretize ranges of items and help interpretation of relative position from the top of a map. However ranking surrounds results of some disadvantages, for instance only visualize properly context of top-100 items, or avoid interactions of items. Complementary methods such as graph visualization should be useful for that. We showed that there exists in some datasets a carefully chosen profile that appeared to indicate something already known about the data. If large amount of visualizations are presented from data to a user, a metrics is probably required to select group of resembling or most salient profiles. Moreover the promised dealing with multivariate data only consists of some hints in high-throughput bio-molecular data.

ACKNOWLEDGMENTS

This work was supported by the French National Institute for Agricultural Research (I.N.R.A.) Grant No. 1077.